\newif\ifsingle
\newif\ifproofs
\documentclass[conference]{IEEEtran}

\usepackage{times}
\usepackage{amsmath,dsfont}
\usepackage{amssymb,amsthm, bm}
\usepackage{epsfig,verbatim}
\usepackage{subfigure}
\usepackage{setspace}
\usepackage{color, soul}
\usepackage{cite}
\usepackage{epstopdf}
\usepackage{graphics}
\usepackage{accents}
\usepackage{acronym}
\usepackage{booktabs}
\usepackage{mathtools}
\usepackage{algorithm}
\usepackage{algorithmic}

\newcommand{\myVec}[1]{{\boldsymbol{#1}}}
\newcommand{\myMat}[1]{{\boldsymbol{#1}}}
\newcommand{\mySet}[1]{\mathcal{#1}}
 \newcommand{\Nr}{N_r}
 \newcommand{\Nu}{N_u}
 \newcommand{\Ntaps}{T}
 \newcommand{\Nclust}{C}
 \newcommand{\Nsubpath}{M_c}
 \newcommand{\Nuset}{\mySet{N}_u}
\newcommand{\myX}{x}			 		
\newcommand{\myY}{{\myVec{y}}}			 		
\newcommand{\myW}{{\myVec{w}}}			 		
\newcommand{\myH}{{\myVec{h}}}	
\newcommand{\WindowSize}{W}
\newcommand{\ModSize}{M}
\newcommand{\MsgSize}{L}

\definecolor{NewColor}{rgb}{0,0,0} %{0.2,0,0.5}

\ifsingle

\else

\fi % ------------------------------------------

\acrodef{adc}[ADC]{analog-to-digital convertor}
\acrodef{cs}[CS]{compressed sensing}
\acrodef{dtft}[DTFT]{discrete-time Fourier transform}
\acrodef{dnn}[DNN]{deep neural network}
\acrodef{ml}[ML]{machine learning}
\acrodef{mmw}[mmWave]{millimeter wave}
\acrodef{rnn}[RNN]{recurrent neural network}
\acrodef{brnn}[BRNN]{bidirectional \ac{rnn}}
\acrodef{csi}[CSI]{channel state information}
\acrodef{map}[MAP]{maximum a-posteriori probability}
\acrodef{snr}[SNR]{signal-to-noise ratio}
\acrodef{bs}[BS]{base station} 
\acrodef{mimo}[MIMO]{multiple-input multiple-output}
\acrodef{mse}[MSE]{mean-squared error}
\acrodef{pdf}[PDF]{probability density function}
\acrodef{pmf}[PMF]{probability mass function}
\acrodef{rv}[RV]{random variable}
\acrodef{lti}[LTI]{linear time-invariant}
\acrodef{wss}[WSS]{wide-sense stationary}
\acrodef{psd}[PSD]{power spectral density}
\acrodef{ser}[SER]{symbol error rate} 
\acrodef{isi}[ISI]{intersymbol interference} 
\acrodef{lstm}[LSTM]{long short-term memory} 
\acrodef{ut}[UT]{user terminal}
\acrodef{gru}[GRU]{gated recurrent unit}
\acrodefplural{gru}[GRU]{gated recurrent units}
\acrodef{em}[EM]{expectation minimization} 
\acrodef{awgn}[AWGN]{additive white Gaussian noise}
\acrodef{cir}[CIR]{channel impulse response}

\IEEEoverridecommandlockouts 

\title{Deep Neural Network Symbol Detection for \\ Millimeter Wave Communications
}
\author{
	\IEEEauthorblockN{Yun Liao\IEEEauthorrefmark{1}, Nariman Farsad\IEEEauthorrefmark{1}, Nir Shlezinger\IEEEauthorrefmark{2}, Yonina C. Eldar\IEEEauthorrefmark{2}, and Andrea J. Goldsmith\IEEEauthorrefmark{1}}
	\IEEEauthorblockA{\\ \IEEEauthorrefmark{1} Department of Electrical Engineering, Stanford University, Stanford, CA, USA \\ \IEEEauthorrefmark{2} Faculty of Mathematics and Computer Science, Weizmann Institute of Science, Rehovot, Israel}
	\thanks{This work was supported in part by the  US - Israel Binational Science Foundation	under grant No. 2026094,  by the Israel Science Foundation under grant No. 0100101, and by the Office of the Naval Research under grant No. 18-1-2191.
		}		
}

\begin{document}
	
	\maketitle
	%\pagestyle{empty}
	%\thispagestyle{empty}
	%----------------------------------------------------------------------------------------
	%	ABSTRACT
	%----------------------------------------------------------------------------------------
	\begin{abstract}
	 This paper proposes to use a deep neural network (DNN)-based symbol detector for mmWave systems such that CSI acquisition can be bypassed. In particular, we consider a sliding bidirectional recurrent neural network~(BRNN) architecture that is suitable for the long memory length of typical mmWave channels. The performance of the DNN detector is evaluated in comparison to that of the Viterbi detector. The results show that the performance of the DNN detector is close to that of the optimal Viterbi detector with perfect CSI, and that it outperforms the Viterbi algorithm with CSI estimation error. Further experiments show that the DNN detector is robust to a wide range of noise levels and varying channel conditions, and that a pretrained detector can be reliably applied to  different mmWave channel realizations with minimal overhead. 
	 
	\end{abstract}

	%----------------------------------------------------------------------------------------
	%	Introduction
	%----------------------------------------------------------------------------------------
	%\vspace{-0.4cm}
	\section{Introduction}
	%\vspace{-0.1cm}
	\iffalse
	TODO Yun and Nir. I recommend as follows:
	\begin{itemize}
		\item Paragraph on motivation for using \acp{dnn} for symbol detection.
		\item Paragraph on millimeter wave communications and its challenges (from a high level perspective).
		\item Paragraph surveying previous works on \ac{ml} for symbol detection and channel decoding. You can base that on Nariman's work \cite{Farsad:18} and the ViterbiNet paper \cite{Shlezinger:19a}. Emphasize what is missing in the previous works in the context of millimeter wave communications.
		\item In this work paragraph, where you explain what we propose in this work.
		\item Paragraph presenting a glimpse into the main results (including numerical results) of this work.
	\end{itemize}
	
	\textcolor{red}{Do not have the page limitation on your mind. Once we have a paper written, we can usually compress it into five pages without removing much content.}
	
	\fi
	
	As one of the important techniques being considered for next-generation wireless communications, communication systems designed to operate in the \ac{mmw} frequency bands have attracted significant attention within the academic and industrial communities. These \ac{mmw} frequency bands offer less congestion than the conventional sub-6GHz bands, while providing multi-gigahertz channel bandwidths with commensurately high data rates \cite{rappaport:13}. In order for \ac{mmw} receivers to reliably decode the transmitted signals, it is critical for them to have accurate knowledge of the underlying channel conditions, namely, to obtain \ac{csi}. However, for \ac{mmw} communications, acquiring accurate \ac{csi} can induce significant overhead as \ac{mmw} channels tend to have very long memory \cite{akdeniz:14}. In particular, this overhead grows dramatically when the communicating nodes are equipped with large antenna arrays, as is often deployed in \ac{mmw} systems to mitigate the large path loss at these frequencies. 
	
	Recently, \acp{dnn} have shown great potential in addressing data detection in wireless channels due to their ability to disentangle the complicated relationship between the channel inputs and outputs in a data-driven fashion, i.e., without explicitly using a channel model. For example, in \cite{nachmani:16}, deep learning was used in decoding linear codes. An end-to-end communication system was trained to jointly optimize encoding and decoding in memoryless channels in \cite{oshea:16}. We note that such an end-to-end approach, in which the channel is treated as an intermediate layer, does not easily extend to mulit-user networks and to channels with memory, as encountered in practical wireless setups. The work \cite{Shlezinger:19a} used \acp{dnn} to learn the weights in the Viterbi algorithm. The computational complexity required to learn these unknown channel conditions using the DNN proposed in \cite{Shlezinger:19a} grows exponentially with the channel memory, thus limiting the feasibility of this technique for equalization of channels with long memory. In \cite{Farsad:18}, the authors proposed a symbol detector based on \acp{rnn}, and in particular, on a sliding \ac{brnn} architecture, for scalar channels, focusing on  molecular  communication with a relatively low data rate. For \ac{mmw} communications,  \acp{dnn} have  been utilized in non decoding-related applications. In \cite{he:2018}, the authors focused on estimating the key characteristics of \ac{mmw} channels with machine learning. The work \cite{huang:2019} used \ac{dnn}s to find hybrid precoding schemes for \ac{mmw} massive MIMO. In \cite{zhou:2019}, the beam management problem in dense \ac{mmw} networks was addressed by deep learning methods.
	
	In this work, we consider using \acp{dnn} for symbol detection in \ac{mmw} communications, bypassing the need to explicitly maintain a channel model with accurate \ac{csi}. In particular, our deep symbol detector extends the sliding \ac{brnn} architecture proposed in \cite{Farsad:18} to \ac{mimo} channels and adapts it to \ac{mmw} systems. We demonstrate that using this \ac{dnn} architecture, it is possible to train a detector that can accurately retrieve the symbols without any prior knowledge of the channel. We also show that the \ac{dnn} decoder generalizes well to different \ac{snr} levels and to small variations in the \ac{cir}. In particular, the sliding \ac{brnn} detector demonstrates improved resiliency to inaccurate training compared to \ac{csi}-dependent detectors such as the Viterbi detector with the same level of \ac{csi} inaccuracy.

	% Organization paragraph
 	The rest of this paper is organized as follows: In Section~\ref{sec:Problem} we discuss the \ac{mmw} channel model and formulate the symbol detection problem. Section~\ref{sec:DNN}  presents the proposed \ac{dnn}-based symbol detector. Section~\ref{sec:Sims} details numerical training and performance results of the proposed detector, and Section~\ref{sec:Conclusions} provides concluding remarks.

 	% Notations paragraph
 	Throughout the paper, we use boldface lower-case letters to denote vectors, e.g., ${\myVec{x}}$;
 	the $i$th element of ${\myVec{x}}$ is written as $\left( {\myVec{x}}\right) _i$. 
% 	The \ac{pdf} of an \ac{rv} $X$ evaluated at $x$ is denoted $\Pdf{X}(x)$,    
Sets are denoted by calligraphic letters, e.g., $\mySet{X}$, and specifically, 
 	$\mySet{Z}$ is the set of integers and 
 	 $\mySet{C}$ is the set of complex numbers.
% 	All logarithms are taken to basis 2. 
% 	Finally, for any sequence, possibly multivariate, $\myVec{y}[i]$, $i \in \mySet{Z}$, and integers $b_1 < b_2$,  $\myVec{y}_{b_1}^{b_2}$ is the column vector $\left[\myVec{y} ^T[b_1],\ldots, \myVec{y}^T[b_2] \right]^T$ and $\myVec{y}^{b_2} \equiv \myVec{y}_{1}^{b_2}$.

	%----------------------------------------------------------------------------------------
	%	Problem Formulation
	%----------------------------------------------------------------------------------------
	%\vspace{-0.2cm}
	\section{Problem Formulation}
	\label{sec:Problem}
	%\vspace{-0.1cm}
	 In order to formulate the \ac{mmw} communications setup, we first describe the overall wireless communication system model in Subsection \ref{subsec:SysModel}, and then elaborate on the specific \ac{mmw} channel model in Subsection \ref{subsec:ChModel}.

	%-----------------------------------
	%	System Model
	%-----------------------------------
	  \subsection{System Model}
	  \label{subsec:SysModel}
	  
	  \begin{figure}[t]
	  \centering
	  \includegraphics[width=0.9\linewidth]{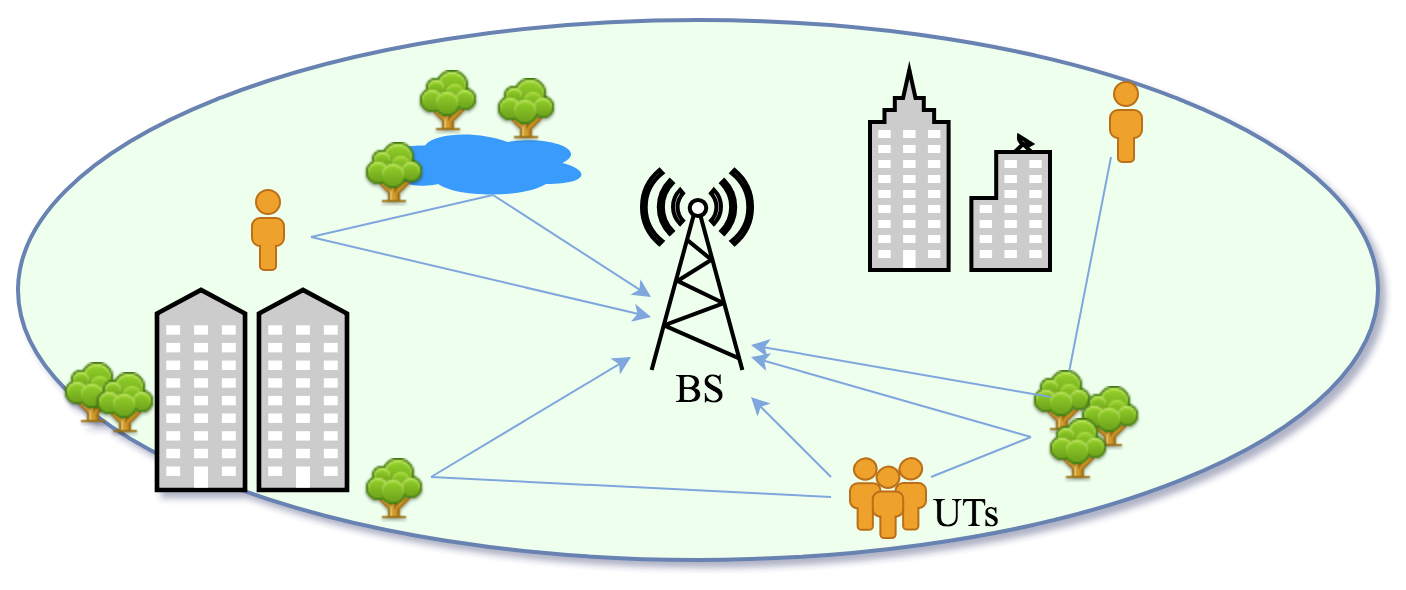}
	  \caption{Wireless communication system illustration.} 
	  \label{fig:system_model}
	  \end{figure}
	  
	  We consider a single-cell multi-user uplink narrowband \ac{mmw} network, in which a \ac{bs} equipped with $\Nr$ antennas serves $\Nu$ single-antenna \acp{ut} as illustrated in Fig.~\ref{fig:system_model}. 
	  The signal transmitted by the $k$th $\ac{ut}$, $k \in \{1,2,\ldots,\Nu\} \triangleq \Nuset$, propagates through a multipath channel with $\Ntaps$ taps, represented by the vectors $\myH_k[l] \in \mySet{C}^{\Nr}$, where $l \in \{0,1,\ldots,\Ntaps-1\}$. 
	  We use $\myW[i] \in \mySet{C}^{\Nr}$ to denote the additive noise at time instance $i$, modeled as an i.i.d. process of zero-mean proper-complex Gaussian random vectors with covariance matrix $\sigma^2 \myMat{I}_{\Nr}$. 
	  Let $\myX_k[i]$ be the signal transmitted by the $k$th user at time instance $i$, assumed to be a digitally modulated symbol of alphabet size $\ModSize > 1$. The channel output observed by the \ac{bs}, denoted $\myY[i]\in\mySet{C}^{\Nr}$, can be written as 
	  \begin{equation}
	  \label{eqn:ChIO}
	      \myY[i] = \sum\limits_{k=1}^{\Nu}\sum\limits_{l=0}^{\Ntaps-1}\myH_k[l]\myX_k[i-l] + \myW[i].
	  \end{equation}
	  
	  Our proposed \ac{dnn}-based receiver recovers the transmitted symbols $\{\myX_k[i]\}$ from the channel outputs $\{\myY[i]\}$. This receiver, detailed in Section \ref{sec:DNN}, is designed to detect the symbols transmitted by a single \ac{ut}, thus implementing {\em separate decoding}, which is a common detection strategy in large-scale \ac{mimo} communications \cite{Marzetta:10}. We leave the implementation of a deep joint detection mechanism, which is known to be superior to separate decoding for such multiple access channels \cite{Verdu:98}, for future investigation.
	  
	  Before presenting the proposed receiver architecture, we first elaborate on the channel model, namely, the model of the vectors $\{\myH_k[l]\}$ which arises in \ac{mmw} communications, in the following subsection. 

	  \subsection{Millimeter Wave Channel Model}
	  \label{subsec:ChModel}
	  %\textcolor{red}{I think you should start this section with briefly mentioning existing \ac{mmw} models, and explain why we adpot the model of \cite{Samimi:16}}. 
	  
	  We adopt the NYU omnidirectional geometric \ac{mmw} channel model proposed in \cite{Samimi:16}. Another widely used \ac{mmw} channel model is given in \cite{3gpptr38901}. The main difference between these two models is in the choice of some of the parameters, so that our proposed design can also be applied under the model of \cite{3gpptr38901}. 
	  
	  The channel model given in \cite{Samimi:16} has a sparse time cluster structure: a \ac{mmw} channel typically exhibits no more than 6 time clusters, with relatively little delay/angle spreading within each cluster. To formulate the model, let $\Nclust$ be the number of time clusters. The $c$-th time cluster contains $\Nsubpath$ subpaths with gains $\{a_{m,c,n}^2\}$, phases $\{\phi_{m,c,n}\}$, and delays $\{\tau_{m,c,n}\}$, $m = 1,2,\ldots, M_c$, $n = 1,2,\ldots, \Nr$. For each $n\in\{1,2,\ldots,\Nr\}$, the $n$th entry of the resulting channel vector $\myH[l]$ can be written as
	  \begin{equation}
	  \label{eqn:ChModel}
	     \left(\myH[l]\right)_n = \sum_{c = 1}^{\Nclust} \sum_{m = 1}^{\Nsubpath} a_{m,c,n} e^{j \phi_{m,c,n}} p(lT_s - \tau_{m,c,n}),
	  \end{equation}
	  where $p(\tau)$ represents the pulse shaping function evaluated at time $\tau$, and $T_s$ is the time interval between two symbols. 
	  
	 \iffalse
	 \textcolor{red}{Is $B$ the bandwidth? Is $T_s$ the sampling period? Do they have to be related via  $T_s=1/B$ or does \eqref{eqn:ChModel} holds for any channel model? And how come the gains / phases / delays do not depend on the antenna index $n$ (which you originally denoted $n_r$?  }
	 \fi
	  The resulting channel tends to exhibit a very long memory, typically several hundred taps. This makes maximum likelihood based algorithms, as well the Viterbi detection algorithm, prohibitively complicated, and, since the \ac{csi} acquisition is more likely to be imprecise in this case when limited channel estimation overhead is allowed, these detection algorithms are more prone to errors. In this sense, an efficient data-driven symbol detection scheme, as proposed in the following section, is highly desirable.
	  
	  \iffalse
	  Here we should detail the millimeter wave channel models, introducing the mathematical notations.
	  Explain that this is a multi-user uplink setup, and illustrate that we propose to utilize a \ac{dnn} for recovering the message of each user separately, leaving a joint deep detector for future investigation. 
	   It would be great to have a figure illustrating the considered input-output relationship. Do not forget to discuss the unique characteristics of such channels. Conclude with the need to implement efficient data-driven symbol detectors for the considered channel, paving the way for the next section.
	   \fi

	%----------------------------------------------------------------------------------------
	%	Deep Symbol Detection
	%----------------------------------------------------------------------------------------
	%\vspace{-0.2cm}
	\section{Deep Symbol Detection}
	\label{sec:DNN}
	%\vspace{-0.1cm}
	In this section, we present the architecture of the proposed \ac{dnn} detector for recovering uplink messages.
	
	%-----------------------------------
	%	Architecture
	%-----------------------------------
	\subsection{DNN Architecture}
	\label{subsec:System}
	\vspace{-0.1cm}
	 \iffalse
	 First elaborate on the model, emphasize in what sense is it different from \cite{Farsad:18}. Adding a diagram of the \ac{dnn} architecture can be helpful.
	 \fi
	 \iffalse
	 \textcolor{red}{(We may want to move this paragraph to the Introduction section or the discussion subsection III.B. I feel it might be better to discuss the reason why we use sliding BRNN before describing the detailed architecture).}
	 \fi
	 Our proposed \ac{dnn} architecture is based on \acp{rnn}, and specifically on an extension of the sliding \ac{brnn} strategy considered in \cite{Farsad:18}. The motivation for utilizing this structure stems from the fact that \acp{rnn} are known to be capable of exploiting temporal correlation among time sequences. However, conventional \acp{rnn} only take the current input and the state, which represents the history, to make predictions. This can be far from optimal in \ac{mmw} symbol detection where the first tap of the \ac{cir} may not be dominant compared to the subsequent taps, especially in non line of sight scenarios. A \ac{brnn}, which takes information from both the history and the future into account in making predictions, helps to improve the performance in this case. The \ac{brnn} scheme, however, also has some drawbacks: the detection algorithm needs to wait until the entire signal is received. To overcome this issue, we adopt the sliding \ac{brnn} architecture, proposed in \cite{Farsad:18} for scalar molecular channels, for symbol detection in multiple-antenna \ac{mmw} communications.
	 
	 \begin{figure}[t]
	  \centering
	  \includegraphics[width=0.9\linewidth]{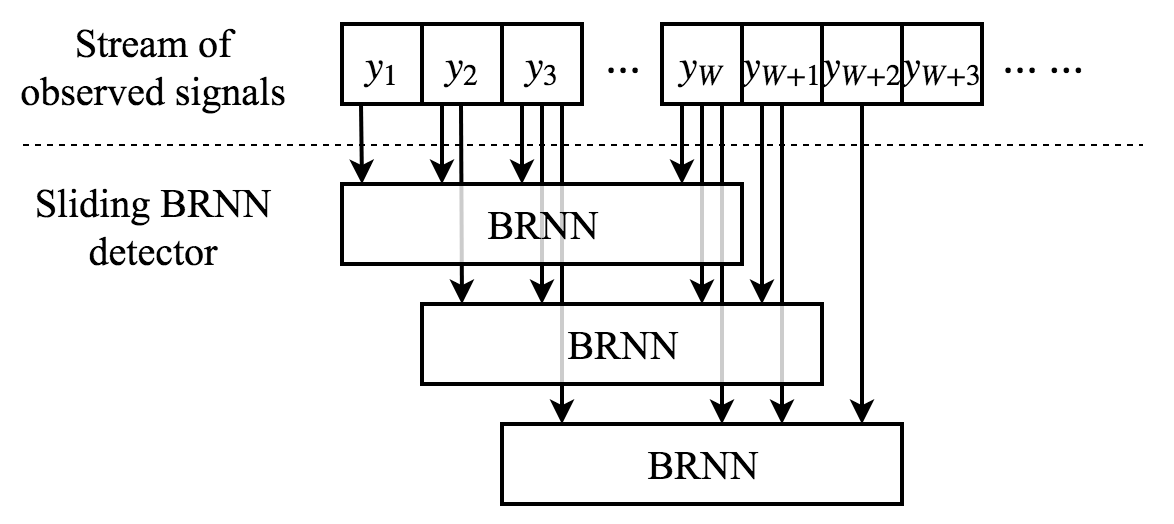}
	  \caption{The sliding BRNN detector.}
	  \label{sbrnn_outer}
	  \end{figure}
	 
	 The detection scheme for recovering the symbols of a single \ac{ut} is illustrated in Fig.~\ref{sbrnn_outer}. When the first $\WindowSize$ symbols arrive at the \ac{bs}, the \ac{brnn} takes the $N_r$ length-$\WindowSize$ sequences from all receive antennas and produces a soft prediction on the symbols $(x_1 \ldots x_{\WindowSize})$ sent by the \ac{ut}. When the next symbol arrives, the \ac{brnn} slides one symbol ahead, and returns the soft prediction on $(x_2, \ldots, x_{\WindowSize+1})$. The procedure is repeated until the entire signal is received. Let  $\mathcal{J}_k = \{j|k \in [j, j+\WindowSize-1]\}$ be the set of all valid starting positions of the sliding \ac{brnn} such that the prediction range covers $x_k$. Let $\bm{p}_k^{(j)}$ be the estimated \ac{pmf} for $x_k$ when the starting position of the \ac{brnn} is $j$.  Note that each $\bm{p}_k^{(j)}$ is an $\ModSize \times 1$ vector. These initial \acp{pmf} are used to obtain the final \ac{pmf} of $x_k$, which is estimated by averaging over $\mathcal{J}_k$, i.e.,
	 \begin{equation}
	     \bm{p}_k = \frac{1}{|\mathcal{J}_k|}\sum_{j \in \mathcal{J}_k} \bm{p}_k^{(j)}.
	 \end{equation}
	 
\begin{figure}[t]
	  \centering
	  \includegraphics[width=\linewidth]{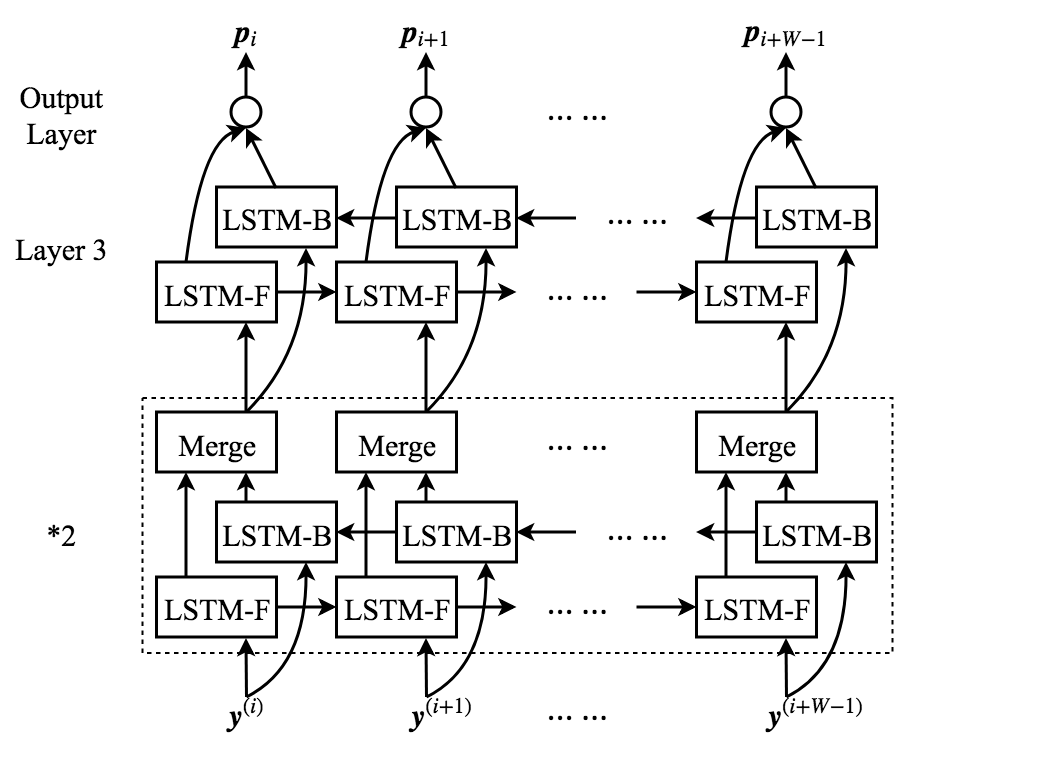}
	  \caption{Architecture of each \ac{brnn}.}
	  \label{sbrnn_inner}
	  \end{figure}

	 Fig.~\ref{sbrnn_inner} shows the architecture of each \ac{brnn} adopted in this work. In particular, \ac{lstm} units with normalized outputs are used as the basic units in the \acp{brnn}. The input $\bm{y}^{(i)} = (y_1^{(i)}, y_2^{(i)}, \ldots, y_{N_r}^{(i)})$ to each \ac{lstm} in the first layer is a $N_r \times 1$ vector containing the received signal at time step $i$ at all receive antennas. Then, three layers of the bidirectional \ac{lstm}s are stacked. The design of the output layer depends on the modulation scheme. When a modulation scheme with $\ModSize>2$ symbols is used by the \acp{ut}, the output layer has output size $\ModSize$ with softmax activation to estimate the \ac{pmf} of the symbols. When a binary constellation, i.e., BPSK, is adopted, we simplify the output layer by using the sigmoid activation function with a scalar output, which represents the estimate of the probability that the symbol is the BSPK symbol $+1$.
	 
	Note that in this work, we do not include channel coding schemes that help recover the original message. The focus of this considered receiver architecture is the reliable detection of the received symbols over \ac{mmw} channels. Since the detector  estimates the \ac{pmf} of the symbol sequence, it is straightforward to add channel coding at the transmitter and then feed the estimated \ac{pmf} to the associated soft channel decoding algorithms in the receiver.
	
	%-----------------------------------
	%	Discussion
	%-----------------------------------
	%\vspace{-0.2cm}
	\subsection{Discussion}
	\label{subsec:Discussion}
	%\vspace{-0.1cm}
	\iffalse
	\textcolor{green}{Here you need to discuss the expected pros and cons of the proposed receiver. 
	Explain why its architecture is suitable for millimeter wave \acp{bs}. 
	Explain in what sense is it different from the deep detector of \cite{Farsad:18}. 
	Explain that we propose to use a dedicated network per each user thus implementing separate decoding, which is a common approach in massive MIMO communications (you can cite here \cite{Marzetta:10} and \cite{Hoydis:13}. You can also say that we consider the implementation of joint-decoding scheme as a topic for future investigation. }
	
	\textcolor{red}{Maybe the first paragraph of III.A is exactly what you expect here?}
	\fi
	
	The proposed sliding \ac{brnn} detector has several main advantages over model-based detectors for \ac{mmw} systems. First, being a data-driven detector, it requires no prior \ac{csi} at the \ac{bs}, thus avoiding the difficult task of accurately estimating the exact \ac{mmw} \ac{cir}. Since the detector learns the channel conditions from the training data, there is no need to do channel estimation separately. As we show in the numerical results in Section \ref{sec:Sims}, the sliding \ac{brnn} detector is quite robust to inaccurate training. In particular, it is demonstrated that a sliding \ac{brnn} detector trained using samples acquired from a set of channel realizations generated according to \eqref{eqn:ChModel} generalizes well to a different \ac{mmw} channel realization.
	
	Furthermore, the sliding detector structure with a moderate window length also allows for near real-time detection. 
	In particular, the proposed detector is capable of recovering the symbol transmitted at time index $i$ once the channel output $\myVec{y}[i+\WindowSize-1]$ is received, i.e., a decoding delay of merely $\WindowSize-1$ samples. To see this, we note that after receiving $\WindowSize-1$ consecutive channel outputs, the detector starts to provide the \ac{pmf} estimations up to the current symbol as soon as the $\WindowSize$th related channel output is obtained.   This allows the detector to recover the symbols  without having to wait for the entire block of channel outputs to be received.
	
	\begin{table}[t]
       \centering
       \caption{Running time comparison.}
       \begin{tabular}{|c|c|c|}
        \hline
       $\Nr$ &  Sliding \ac{brnn} &  Viterbi \\ \hline
       4 & {\bf 0.244} s & 12.461 s \\ \hline
       128 & {\bf 0.264} s & 52.681 s \\ \hline
       \end{tabular}
       \label{tab:complexity}
   \end{table}
   
   Finally, a notable advantage of the sliding \ac{brnn} architecture compared to model-based approaches stems from the fact that, once trained, the decoding can be carried out significantly faster than iterative model-based approaches, and that this delay hardly increases as the number of antennas $\Nr$ grows. This gain is illustrated Table \ref{tab:complexity}, which shows the average running time of detecting one 200-bit message using the sliding \ac{brnn} detector and the Viterbi detector measured on an AMD Phenom(tm) II X6 1045T 2.70 GHz Processor. This makes the sliding \ac{brnn} a promising decoding approach for \ac{mmw} receivers, which typically utilize massive \ac{mimo} arrays.

	\section{Numerical Study}
	\label{sec:Sims}
	%\vspace{-0.1cm}
	%\textcolor{green}{TODO Yun and Nir}
	
   %\textcolor{green}{Since I assume that this section will be long, I also recommend to divide it into subsections, including:}
   
   In this section, we numerically evaluate the performance of our \ac{dnn}-based symbol detector in terms of the detection accuracy and the convergence speed in training. The main characteristics of the simulated environment in this section are listed in Table \ref{tab:params}. The parameters $C$, $M_c$, $\{a_{m,c,n}\}$, $\{\phi_{m,c,n}\}$, $\tau_{m,c,n}$ in \eqref{eqn:ChModel} are generated according to the model suggested in \cite{Samimi:16} under the parameter settings of Table  \ref{tab:params}. For the sliding \ac{brnn} detector, the hidden size of each \ac{lstm} unit is 20, and the window length $\WindowSize$ is set to 30. For the Viterbi detector, we use beam-search \cite{Lingyun:04} with $300$ survivor paths.
   
   \begin{table}[t]
       \centering
       \caption{Key parameters in simulations.}
       \begin{tabular}{c|c}
        \hline
       {\bf Parameter} & {\bf Value} \\ \hline
       Carrier Frequency & 28 GHz \\ \hline
       Bandwidth & 800 MHz \\ \hline
       Cell Type & urban microcell \\ \hline
       Antenna array & uniform linear array \\ \hline
       Antenna spacing & 0.5 wavelength \\ \hline
       Transmit power & 11 dBm \\ \hline
       Tx-Rx Separation distance & 60 m \\ \hline
       Tx-Rx Antenna Gains & 24.5 dBi \\ \hline
       \end{tabular}
       \label{tab:params}
   \end{table}

   	%-----------------------------------
	%	Error Rate Comparison
	%-----------------------------------
	\subsection{Error Rate Comparison}
	\label{subsec:ErrorRate}
	%\vspace{-0.1cm}	
	%\textcolor{green}{This is the main simulations study where you compute error rates and evaluate robustness to \ac{csi} uncertainty. }
	
	In this subsection, the performance of the sliding \ac{brnn} detector is compared with that of the Viterbi algorithm with either perfect or imperfect \ac{csi} in terms of \ac{ser}. In the simulations, BPSK modulation is used. The number of receive antennas is fixed to $\Nr = 4$. For each \ac{snr} value, the \ac{ser} is averaged over 5 different independently generated \ac{mmw} channel realizations, and $2 \times 10^5$ symbols, divided into blocks of $200$ symbols, are detected for each channel realization. 
	
	\begin{figure}
	    \centering
	    \includegraphics[width=0.9\linewidth]{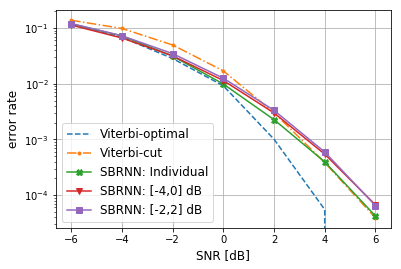}
	    \caption{Comparison of \ac{ser} between the sliding \ac{brnn} detector and Viterbi with perfect \ac{csi}. The sliding \ac{brnn} detector is trained with or without \ac{snr} uncertainty.}
	    \label{fig:ber_curve_snr_uncertainty}
	\end{figure}
	
	We first evaluate the \ac{ser} performance of  the sliding \ac{brnn} detector compared to the Viterbi detector with perfect \ac{csi}. The resulting \ac{ser} values versus \ac{snr} are depicted in Fig.~\ref{fig:ber_curve_snr_uncertainty}. 
	The dashed curve in Fig.~\ref{fig:ber_curve_snr_uncertainty} is the error rate of the optimal Viterbi detector in which the detector has perfect \ac{csi} and waits to receive the entire block with the tail before detecting the symbols. This receiver implements maximum-likelihood detection for the \ac{mmw} channel. Since the sliding \ac{brnn} detector gives the detection results as soon as it receives the last symbol in the message, we also evaluate the performance of the Viterbi detector that only considers a finite window of the received signal, whose window size equals the message length.  The performance of the Viterbi detector in this case is shown by the dot dashed curve with the legend ``Viterbi-cut''. 
	The solid curve with cross markers in Fig.~\ref{fig:ber_curve_snr_uncertainty} shows the error rate of the sliding \ac{brnn} detector that is trained and evaluated at the same \ac{snr} level under the same channel realization. The training set for each \ac{snr} contains 4,000 training samples, and each training sample is a tuple of a 200-symbol transmitted block $\{x[i]\}$ and the corresponding channel outputs $\{\myVec{y}[i]\}$. It is clear that the performance of the sliding \ac{brnn} detector is close to that of the optimal Viterbi detector when the \ac{snr} is low. Note that the sliding \ac{brnn} detector only takes the received signal in a limited window for prediction, while the optimal Viterbi detector utilizes the entire received signal. When taking in the same amount of signal, the sliding \ac{brnn} detector outperforms the Viterbi-cut detector over a large range of \ac{snr} levels. This demonstrates the ability of the data-driven sliding \ac{brnn} receiver to reliably detect the transmitted symbols without requiring prior knowledge of the \ac{csi}.
	
	To represent practical setups in which the exact value of the \ac{snr} is not known during training, we also depict in Fig.~\ref{fig:ber_curve_snr_uncertainty} the \ac{ser} performance of the sliding \ac{brnn} receiver when trained using training samples corresponding to several noise levels. The legends of the curves in Fig.~\ref{fig:ber_curve_snr_uncertainty} show the range of \ac{snr} in the training signal streams. In particular, to generate the training set, we randomly pick an \ac{snr} value uniformly from the given range as the \ac{snr} for each block. After training, we use the same sliding \ac{brnn} to detect signal streams under different \ac{snr} levels. As shown in Fig.~\ref{fig:ber_curve_snr_uncertainty}, the performance of the sliding \ac{brnn} decoders remain within a small gap from the Viterbi detector. 
	It is also interesting that the sliding \acp{brnn} trained across different \ac{snr} ranges yield similar performance. Evidently, the trained sliding \ac{brnn} generalizes well to a wide range of \acp{snr}. These observations indicate that it suffices to train a sliding \ac{brnn} under a single \ac{snr} and use it regardless of the change to the \ac{snr}, resulting in a robust symbol detection scheme in time-varying \ac{snr} environments.
	
	\begin{figure}
	    \centering
	    \includegraphics[width=0.9\linewidth]{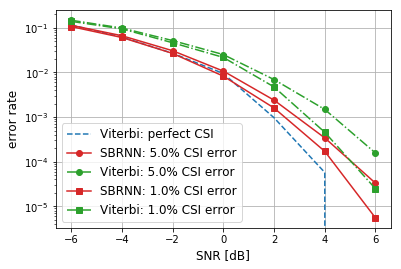}
	    \caption{Comparison of \ac{ser} between sliding \ac{brnn} detector and Viterbi with imperfect \ac{csi}.}
	    \label{fig:ber_curve_csi_uncertainty}
	\end{figure}

	Next, we numerically evaluate the generalization capability of the sliding \ac{brnn} under \ac{csi} uncertainty. 
	In particular,  we consider the case where the \ac{csi} estimate is imprecise or the channel varies slightly over time. With this aim, we evaluate the \ac{ser} of the sliding \ac{brnn} detector on a \ac{mmw} channel realization whose \ac{cir} varies by $2.5\%$ or $1.0\%$ compared to the \ac{cir} used during training. Specifically, the sliding \ac{brnn} is trained under one particular channel realization and a 0 dB \ac{snr}. Then, in evaluation, we randomly distort the amplitude at each tap $\{a_{m,c,n}\}$ by $2.5\%$ or $1.0\%$ on average for each message block compared to the channel used during training. For comparison, we use the Viterbi detector with the same level of \ac{csi} error. The results of this simulation are depicted in Fig.~\ref{fig:ber_curve_csi_uncertainty}, where it is observed that the sliding \ac{brnn} detector consistently outperforms the Viterbi detector with imperfect instantaneous \ac{csi}, and that the performance of the sliding \ac{brnn} remains close to the optimal Viterbi detector with perfect \ac{csi}. The result shows that the sliding \ac{brnn} detector is able to generalize well to not only the various noise levels, but also to different channel realizations.
	
	\begin{figure}[t]
	    \centering
	    \includegraphics[width=0.9\linewidth]{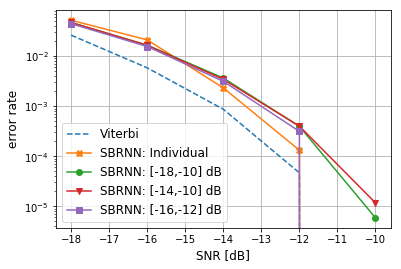}
	    \caption{\ac{ser} of sliding \ac{brnn} detector for SIMO with 128 receive antennas.}
	    \label{fig:ber_curve_massive}
	\end{figure}
	
	The numerical evaluations presented in Figs.~\ref{fig:ber_curve_snr_uncertainty}-\ref{fig:ber_curve_csi_uncertainty} focused on a \ac{bs} with only $4$ receive antennas. In order to demonstrate that the gains of the sliding \ac{brnn} receiver are also maintained for large antenna arrays, which are commonly utilized in \ac{mmw} communications, we depict in Fig.~\ref{fig:ber_curve_massive}  the \ac{ser} curves of the sliding \ac{brnn} detector when the number of receive antennas is $\Nr =128$. While we do not increase the size of the neural network or the window length, the performance of the sliding \ac{brnn} detector is still close to the Viterbi detector, while the running time is significantly reduced, as discussed in Subsection \ref{subsec:Discussion}. Similarly to the results with fewer receive antennas presented in Fig.~\ref{fig:ber_curve_snr_uncertainty}, a single sliding \ac{brnn} detector trained under a wide range of \acp{snr} generalizes well to different noise levels. These results demonstrate the great potential of using a data-driven \ac{dnn}-based detector for \ac{mmw} communications.

	%-----------------------------------
	%	Training Size Analysis
	%-----------------------------------
	\subsection{Training Size Analysis}
	\label{subsec:Training}
	%\vspace{-0.1cm}
	%\textcolor{green}{Here you can present the convergence versus epochs numerical test, demonstrating that the proposed \ac{dnn} requires relatively little time to converge, which indicates its potential ability to train online in the presence of time-varying channel conditions.}

    Next, we numerically study the convergence speed in training the sliding \ac{brnn} detector.
 %  
   \iffalse
   \begin{figure}[t]
       \centering
       \includegraphics[width=0.9\linewidth]{conv_speed_massive_32.png}
       \caption{Convergence speed of \ac{dnn} detector for $1 \times 128$ SIMO case.}
       \label{fig:conv_massive}
   \end{figure}

   Fig.~\ref{fig:conv_massive} shows the convergence speed of the \ac{brnn} in the massive SIMO case. It could be seen that 4,000 blocks of messages are sufficient to train the network from scratch.
   
   \fi
 %  
   \begin{figure}[t]
       \centering
       \includegraphics[width=0.9\linewidth]{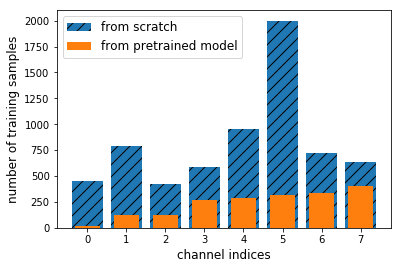}
       \caption{Comparison of convergence speed (number of training samples needed) for the sliding \ac{brnn} detector when trained from an existing model for a completely different channel realization.}
       \label{fig:conv_compare}
   \end{figure}

   To that aim, we depict in  Fig.~\ref{fig:conv_compare} the convergence time of the sliding \ac{brnn} detector. We consider two different initial values for the network weights: in the first case (filled bars in Fig.~\ref{fig:conv_compare}) the detector was already trained under some channel, and needs to be re-trained for a different channel realization. This simulates the practical scenario where the channel changes over time, and the detector needs to be adjusted to track the channel periodically. In the second case (slashed bars in Fig.~\ref{fig:conv_compare}) the weights in the sliding \ac{brnn} are randomly initialized, representing the scenario where the network is trained from scratch. In particular, $9$ different channel realizations are generated independently, indexed $0,\ldots,8$, and a sliding \ac{brnn} symbol detector trained on the last realization is utilized as the pretrained model. For each of the other $8$ channel realizations, we train the sliding \ac{brnn} detector either from scratch or from the pretrained model with the same learning rate. Fig.~\ref{fig:conv_compare} shows the number of training samples required to achieve at least $90\%$ detection accuracy. The channels are sorted by the number of the required samples for training from a pretrained model to illustrate the distribution of the required number. Observing Fig.~\ref{fig:conv_compare}, we note that it takes many fewer samples for the sliding \ac{brnn} detector to be adapted to a new channel compared to training it from scratch. In particular, for some channel realizations, such as channels with indices 0, 1, and 2 in the figure, it takes less than $150$ samples for the detector to adapt to the new channel. This study indicates that a previously trained sliding \ac{brnn} architecture can be adjusted to time-varying channel conditions with minimal overhead for re-training the network.

	%----------------------------------------------------------------------------------------
	%	CONCLUSIONS
	%----------------------------------------------------------------------------------------

	\section{Conclusions}
	\label{sec:Conclusions}
	
	This paper studied a \ac{dnn}-based symbol detector for uplink \ac{mmw} communications, which does not require \ac{csi} and learns the decoding mapping in a data-driven fashion. Our proposed receiver was based on a sliding \ac{brnn} architecture, which is suitable for the long channel memory and large number of receive antennas commonly used in \ac{mmw} systems. Numerical evaluations demonstrated that the sliding \ac{brnn} detector is capable of achieving performance within a small gap from the optimal Viterbi detector, which requires full \ac{csi}. Moreover, the sliding \ac{brnn} detector was shown to generalize well when trained using samples taken from an inaccurate model, and in fact it outperformed the Viterbi detector with the same level of \ac{csi} uncertainty. Finally, we numerically showed that a pretrained \ac{dnn} detector could be adjusted to another channel realization with minimal overhead.

	%----------------------------------------------------------------------------------------
	%	BIBLIOGRAPHY
	%----------------------------------------------------------------------------------------

\end{document}